\documentclass[apj]{emulateapj}

\usepackage{natbib}
\usepackage{color}

\usepackage{bm}
\usepackage{lscape}

\newcommand{\eg}{e.g., }
\newcommand{\ie}{i.e., }
\newcommand{\Msun}{M_{\odot}}
\newcommand{\kms}{km~s$^{-1}$}

\newcommand{\Nifs}{$^{56}$Ni}

\newcommand{\Mej}{M_{\rm ej}}
\newcommand{\KE}{E_{\rm K}}
\def\gsim{\mathrel{\rlap{\lower 4pt \hbox{\hskip 1pt $\sim$}}\raise 1pt
\hbox {$>$}}}
\def\lsim{\mathrel{\rlap{\lower 4pt \hbox{\hskip 1pt $\sim$}}\raise 1pt
\hbox {$<$}}}

\shorttitle{Radioactively Powered Emission from BH-NS Mergers}
\shortauthors{Tanaka et al.}

\begin{document}

\title{
Radioactively Powered Emission from Black Hole-Neutron Star Mergers}
\author{
Masaomi Tanaka\altaffilmark{1},
Kenta Hotokezaka\altaffilmark{2},
Koutarou Kyutoku\altaffilmark{3},
Shinya Wanajo\altaffilmark{1},
Kenta Kiuchi\altaffilmark{4}, \\
Yuichiro Sekiguchi\altaffilmark{4}, and
Masaru Shibata\altaffilmark{4}
}

\altaffiltext{1}{National Astronomical Observatory of Japan, Mitaka, Tokyo 181-8588, Japan; masaomi.tanaka@nao.ac.jp}
\altaffiltext{2}{Department of Physics, Kyoto University, Kyoto 606-8502, Japan}
\altaffiltext{3}{Department of Physics, University of Wisconsin-Milwaukee, P.O. Box 413, Milwaukee, Wisconsin 53201, USA}
\altaffiltext{4}{Yukawa Institute for Theoretical Physics, Kyoto University, Kyoto 606-8502, Japan}

\begin{abstract}
Detection of electromagnetic counterparts of gravitational wave (GW) 
sources is important to unveil the nature of compact binary coalescences.
We perform three-dimensional, time-dependent, multi-frequency
radiative transfer simulations for radioactively powered emission 
from the ejecta of black hole (BH) - neutron star (NS) mergers.
Depending on the BH to NS mass ratio, spin of the BH, and 
equations of state of dense matter, 
BH-NS mergers can eject more material than NS-NS mergers.
In such cases, radioactively powered emission from the BH-NS merger ejecta 
can be more luminous than that from NS-NS mergers.
We show that, in spite of the expected
larger distances to BH-NS merger events,
observed brightness of BH-NS mergers can be comparable to
or even higher than that of NS-NS mergers.
We find that,
when the tidally disrupted BH-NS merger ejecta 
are confined to a small solid angle,
the emission from BH-NS merger ejecta
tends to be bluer than that from NS-NS merger ejecta 
for a given total luminosity.
Thanks to this property, we might be able to distinguish 
BH-NS merger events from NS-NS merger events
by multi-band observations of the radioactively powered emission.
In addition to the GW observations,
such electromagnetic observations can 
potentially provide independent information 
on the nature of compact binary coalescences.
\end{abstract}


\section{Introduction}
\label{sec:intro}

Next-generation gravitational wave (GW) detectors, 
such as, Advanced LIGO, Advanced Virgo, and KAGRA 
\citep{abadie10,kuroda10,accadia11,ligo13}, 
are expected to detect GW signals from compact binary coalescences.
Mergers of black hole (BH) and neutron star (NS) binary
are among the promising sources of GWs
\citep[see][for a review]{shibata11}.
Although the BH-NS merger rate is 
estimated to be lower than the NS-NS merger rate by a factor of $\sim 10-100$, 
the detection rates can be comparable thanks to 
expected larger horizon distances
to the BH-NS mergers by a factor of about 2.
The expected detection
rate of BH-NS mergers with the advanced detectors is $\sim$ 10 yr$^{-1}$
(with the lower and upper estimation of 0.2 and 300 yr$^{-1}$, 
respectively, see \citealt{abadie10rate}).

Even if GWs are detected by several detectors, 
the position of the GW source cannot be accurately determined
(a typical localization is 
about 10-100 deg$^2$, \eg \citealt{ligo12,ligo13}).
Therefore, to study the nature of the GW sources,
identification of electromagnetic (EM) counterparts is crucial
\citep{nissanke13,kelley13,kasliwal13}.
Short gamma-ray bursts (GRBs) are promising candidates 
\citep{kochanek93},
but the association fraction will not be large 
if the jet is beamed into a small solid angle.
Motivated by these facts, possible isotropic EM signals from GW sources 
have been suggested
\citep[\eg][]{li98,metzger12,kyutoku12,rosswog13,piran13,takami13}.

The emission powered 
by radioactive energy of $r$-process nuclei 
is one of the important targets for follow-up EM observations.
For the case of NS-NS mergers,
a part of material is expected to be ejected
\citep[\eg][]{rosswog99,lee07,goriely11,hotokezaka13,bauswein13},
and $r$-process nucleosynthesis is thought to take place in the ejecta
\citep[\eg][]{symbalisty82,eichler89,meyer89,freiburghaus99,roberts11,goriely11,korobkin12,bauswein13,rosswog13b}.
By the decay energy of the synthesized nuclei, 
the ejecta can emit ultraviolet-optical-infrared (UVOIR)  
radiation \citep{li98}.
This is a similar emission mechanism to supernovae, 
where \Nifs\ is the dominant heating source.
Such an emission from NS-NS mergers has been called 
``kilonova'', ``macronova'', or ``mini-SN''
\citep{kulkarni05,metzger10,roberts11,goriely11,metzger12}.

\begin{figure*}
\begin{center}
\begin{tabular}{cc}
\includegraphics[scale=1.2]{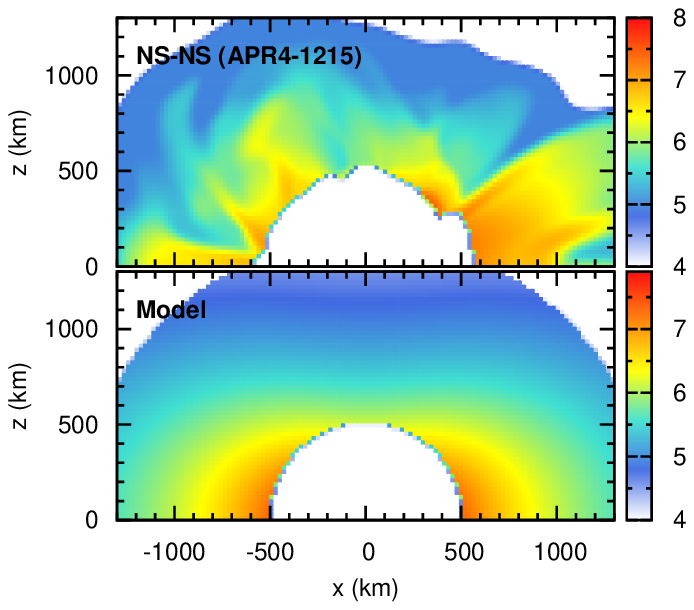} &
\includegraphics[scale=1.2]{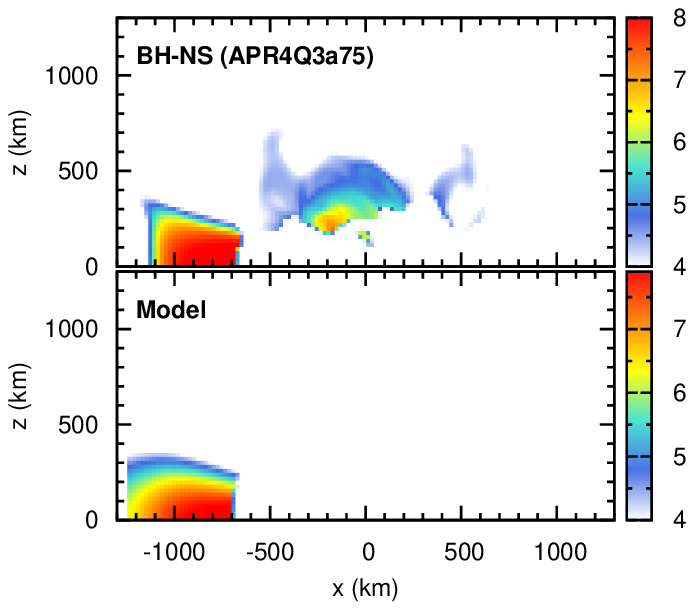} 
\end{tabular}
\caption{
Density distributions of the NS-NS merger model APR4-1215 
\citep{hotokezaka13} (Left)
and the BH-NS merger model APR4Q3a75 \citep{kyutoku13} (Right)
at $t \sim 10 $ ms after the merger.
The top panels show the results of numerical relativity simulations
while the bottom panels show the modeled, smoothed density structures,
which are used as input of radiative transfer simulations.
The vertical ($z$) axis is perpendicular to the orbital plane 
of the merger.
The color bar indicates $\log \rho$ in ${\rm g\ cm^{-3}}$.
In BH-NS mergers, the mass ejection is non-axisymmetric
and concentrated near the equatorial plane.
Only the escaping material is shown in the plots.
}
\label{fig:models}
\end{center}
\end{figure*}

\citet{kasen13}, \citet{barnes13}, and \citet[hereafter TH13]{tanaka13}
performed radiative transfer simulations for radioactively powered 
emission from NS-NS mergers, taking into account 
the wavelength-dependent opacity of $r$-process elements.
They found that (1) the luminosity is about $10^{41} \ {\rm erg \ s^{-1}}$
(for the ejecta mass of $\Mej = 10^{-2} \Msun$),
(2) the spectral energy distribution (SED) peaks at 
the red edge of optical and near-infrared (NIR) wavelengths, 
and (3) the emission lasts about 5-20 days.
In fact, following the {\it Swift} detection of the short GRB 
130603B (\citealt{melandri13}, see also \citealt{deUgartePostigo13}), 
a bright NIR excess was detected in the afterglow 
\citep{berger13,tanvir13}.
This NIR emission can be interpreted as a
radioactively powered emission, being broadly consistent with 
the results of radiative transfer simulations
(\citealt{berger13,tanvir13,hotokezaka13b})
\footnote{See \citet{jin13} for an alternative scenario,
involving the synchrotron radiation by a wide, mildly relativistic outflow,
although non-detection of late-phase radio emission does not 
support this scenario \citep{fong13}.}.
If this interpretation is the case, this discovery 
suggests that radioactively powered emission 
actually takes place in binary coalescences,
and it could be used for the localization of GW sources.

The mass ejection and the $r$-process are 
also expected in BH-NS mergers \citep{lattimer74,lattimer76}.
Depending on the BH to NS mass ratio, spin of the BH, and 
equations of state (EOS) adopted in the merger simulations,
BH-NS mergers can eject more material than NS-NS mergers
\citep{kyutoku11,foucart13,lovelace13,deaton13,kyutoku13}.
\citet{kyutoku13} also showed that the mass ejection can be highly asymmetric.
They discussed impacts of the asymmetric mass ejection 
on the properties of the EM counterparts.

In this paper, we study properties of 
radioactively powered emission from BH-NS mergers.
We perform three-dimensional, 
time-dependent, multi-frequency radiative transfer simulations 
of the BH-NS merger ejecta for the first time.
In Section \ref{sec:models}, 
we describe our models of BH-NS mergers and methods of simulations.
Results of radiative transfer simulations are shown
in Section \ref{sec:results}.
Implications for EM observations following GW detection
are discussed in Section \ref{sec:implications}.
Finally, we give conclusions in Section \ref{sec:conclusions}.

\begin{deluxetable}{lccccc} 
\tablewidth{0pt}
\tablecaption{Summary of Models employed for the radiative transfer simulations}
\tablehead{
Model & $M_{\rm BH}$  & $M_{\rm NS}$ &  $\Mej$   & $\KE$  & $v_{\rm ch}$  \\
(BH-NS) & ($\Msun$)   & ($\Msun$)   & ($\Msun$) & (erg)   &  ($c$)        
}
\startdata
APR4Q3a75 & 4.05 & 1.35 & $1 \times 10^{-2}$    &  $5 \times 10^{50}$  & 0.24  \\
H4Q3a75    & 4.05 & 1.35 & $5 \times 10^{-2}$    &  $2 \times 10^{51}$  & 0.21  \\
MS1Q3a75   & 4.05 & 1.35 & $7 \times 10^{-2}$    &  $4 \times 10^{51}$  & 0.25  \\
\hline \hline \\

Model    & $M_{\rm NS}$  & $M_{\rm NS}$ &  $\Mej$   & $\KE$  & $v_{\rm ch}$  \\
(NS-NS)  & ($\Msun$)   & ($\Msun$)   & ($\Msun$) & (erg)   &  ($c$)      \\
\hline 
APR4-1215  &  1.2   & 1.5     &  $0.9 \times 10^{-2}$  & $4 \times 10^{50}$ & 0.24 \\
H4-1215   &  1.2   & 1.5     &  $0.4 \times 10^{-2}$   & $1 \times 10^{50}$ & 0.21 
\enddata
\label{tab:models}
\end{deluxetable}

\section{Models and Methods}
\label{sec:models}

\subsection{Models}

We use the results of numerical-relativity simulations by 
\citet{kyutoku13} as input models of our radiative transfer simulations.
We adopt three models with different EOSs.
The adopted EOSs are APR4 \citep{akmal98}, 
H4 \citep{glendenning91,lackey06}, and MS1 \citep{muller96}.
APR4 is a ``soft'' EOS, giving the radius of 11.1 km for a 1.35 $\Msun$ NS,
while H4 and MS1 are ``stiff'' EOSs, 
giving the radii of 13.6 km and 14.4 km for a 1.35 $\Msun$ NS, respectively.
For more details on these EOSs, see \citet{hotokezaka13}.
For all the models, the gravitational mass of the NS is fixed to be 
$M_{\rm NS} = 1.35 \Msun$.
The mass ratio of BH to NS, $Q = M_{\rm BH} / M_{\rm NS}$, 
is set to be $3$ ($M_{\rm BH} = 4.05 \Msun$).
In all the models, 
the non-dimensional spin parameter of the BH is set to be $\chi = 0.75$,
with the direction aligned with the 
binary orbital angular momentum

With this BH spin parameter,
NSs are tidally disrupted by the BHs 
with a wide range of the mass ratios and EOSs.
The ejecta masses ($\Mej$) do not depend strongly on the mass ratios
except for the soft EOSs, such as APR4.
For the stiff EOSs H4 and MS1, 
the ejecta masses are not smaller than $10^{-2} \Msun$, 
\ie $\Mej \sim (4 - 8) \times 10^{-2} \Msun$
(for $Q=3-7$ and $\chi = 0.75$, \citealt{kyutoku13}, 
see also Table \ref{tab:models}).
For the soft EOS APR4, the ejecta mass decreases as the BH mass increases,
from $\Mej \sim 1 \times 10^{-2} \Msun$ (for $Q=3$)
down to $\Mej \sim 5 \times 10^{-4} \Msun$
(for $Q=7$, see \citealt{hotokezaka13b}).
The mass ejection is also affected by the BH spin.
When a non-spinning BH is considered, the ejecta mass can be 
substantially smaller than $10^{-2} \Msun$ (K. Kyutoku et al. in preparation).
In this paper, we focus on the three models 
with relatively efficient mass ejection.

As described by \citet{li98}, 
the behavior of radioactively powered emission 
is mainly determined by the mass of the ejecta 
and the characteristic velocity ($v_{\rm ch} = \sqrt{2\KE/\Mej}$, 
where $\KE$ is the kinetic energy of the ejecta).
These parameters are summarized in Table \ref{tab:models}.
Even with the different binary parameters,
the behaviors of the emission are expected to resemble each other 
as long as the ejecta mass and characteristic velocity are similar.
We see further effects by the ejecta geometry in Section \ref{sec:results}.

Figure \ref{fig:models} shows the density distribution of 
the NS-NS merger model APR4-1215 (left) and
the BH-NS merger model APR4Q3a75 (right).
As demonstrated by \citet{kyutoku13}, 
the mass ejection from BH-NS mergers can be highly asymmetric.
We first remap the density distribution of the ejecta 
into a two-dimensional, axisymmetric model, 
imposing north-south symmetry.
The mass ejection found in the numerical-relativity
simulations is, however, not axisymmetric, but occurs to a particular direction
near the equatorial plane with an opening angle of about 180 deg.
To reproduce such anisotropic mass ejection,
we omit this modeled, axisymmetric density structure at $x > 0$,
and enhance the density at $x < 0$ 
by a factor of 2 keeping the total ejecta mass.
Even with this simplification,
our model still captures the global density distribution
which characterizes the BH-NS merger ejecta
(bottom panels of Figure \ref{fig:models}).
The numerical-relativity simulations typically follow
the dynamics of BH-NS coalescences up to about $t=10$ ms after the merger.
On the other hand, we start radiative transfer simulations 
from $t = 0.1$ days.
For the evolution between these epochs, 
we simply scale down the density as $\rho \propto t^{-3}$.
Note that this assumption might overestimate the density 
at later epochs because the continuous radioactive heating is neglected
\citep[see][]{rosswog13b}.

\begin{figure}
\begin{center}
\includegraphics[scale=1.2]{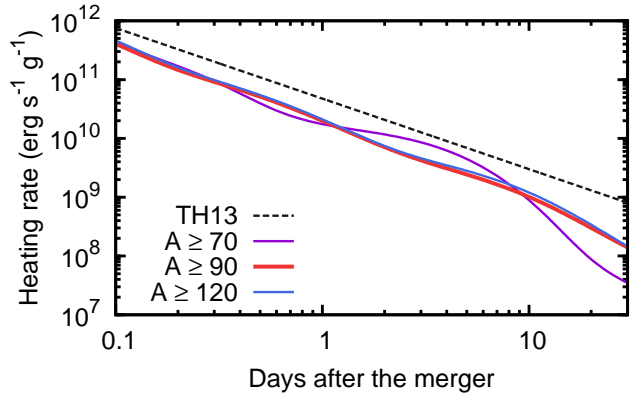}
\caption{
The heating rate from the $\beta$-decays of $r$-process nuclei.
Purple, red and blue lines show the heating rates computed 
from the initial conditions reproducing the 
solar $r$-process abundance ratios with $A \ge 70$, 
$A \ge 90$, and $A \ge 120$ respectively.
The dashed line shows the approximated heating rate adopted by TH13.
In this paper, we use the heating rate with $A \ge 90$ (red line).
}
\label{fig:Qdot}
\end{center}
\end{figure}

\subsection{Methods}

For radiative transfer, 
we use three-dimensional, time-dependent, multi-frequency 
Monte Carlo radiative transfer code developed by TH13.
For a given density structure and abundance distribution,
the code computes the time series of spectra 
in the UVOIR wavelengths.
The code adopts three-dimensional Cartesian grid, 
typically with $32^3$ cells.
Thanks to the nearly homologous expansion,
velocity is used as a spatial coordinate.
For the models presented in this paper, 
a typical spatial resolution is $\Delta v \sim $ 2000 \kms.
For the time grid, we use a logarithmically-spaced time step,
with a time step of $\Delta \log (t/{\rm day}) = 0.05$.
For the frequency grid, we use a linearly-spaced 
grid in the wavelength, 
$\lambda =$ 100 - 25000 \AA\ with $\Delta \lambda = 10$ \AA.

A major update of the code is the
heating rate, for which TH13 adopted only an approximated recipe.  
The radioactive heating rate from $r$-process nuclei has been studied by
several authors according to nucleosynthesis calculations
\citep[\eg][]{metzger10,goriely11,korobkin12,grossman13}.  
\citet{grossman13} pointed out that the heating rate 
computed from their calculations was 
lower than that assumed in TH13.  They also showed that the heating rate
depended on the initial $Y_\mathrm{e}$ (number of protons per nucleon)
in the ejecta. 
It is noted that a prediction of the
$r$-process abundance curve is subject to uncertainties in the
astrophysical models as well as in the theoretical nuclear data adopted in
nucleosynthesis calculations \citep[\eg][]{metzger10}. 

For demonstrative purposes in this paper, 
we simply assume the time-independent abundance distribution as a 
function of $A$, $Y_A$ (number of nuclei with $A$ per nucleon),
being the same as the solar system $r$-process pattern \citep{cowan99}.
Note that $Y_A$ is the sum of the time-dependent abundances
of the isobars with different atomic numbers $Z$
in the neutron-rich side of $\beta$-stability.
We have computed the heating rate starting from the
initial compositions $Y_A$ at the neutron separation energies of 2~MeV
(roughly at the $r$-process freezeout; the result is almost independent
of this value). 
Here, heating is due to $\beta$-decays that do not
change $A$ but increase $Z$ of a given nuclide; 
$\beta$-delayed neutron emission (that changes $A$), 
which plays a role only during the first seconds, are not considered. 
After several seconds, most of the nuclei decay back to
the vicinity of $\beta$-stability, 
where experimental halflives \citep{horiguchi96,nishimura11} 
and $Q$-values (from the nuclear masses compiled by 
G. Audi and W. Meng 2011, private communication) are available. 
The uncertainties originating from the theoretical nuclear
data are thus irrelevant.
Heating from nuclear fission (that changes $A$) is not considered here,
which should be subdominant in our case
(assuming the solar $r$-process pattern of $Y_A$'s)
because of the abundant $A \sim 130$ nuclei
that dominate the radioactive energies \citep{metzger10}.

\begin{figure}
\begin{center}
\includegraphics[scale=1.3]{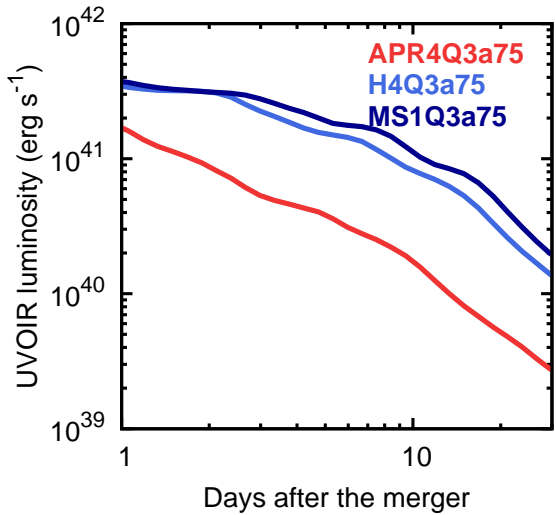}
\caption{
Bolometric light curves of the BH-NS merger models.
The luminosities are those averaged over all solid angles.
Different colors show the models with 
different EOSs adopted in the merger simulations.
The BH-NS mergers with stiff EOSs (H4 and MS1) 
are brighter
than that with a soft EOS (APR4)
because of the larger ejecta mass for the stiffer EOSs.
}
\label{fig:Lbol_BH}
\end{center}
\end{figure}

Figure \ref{fig:Qdot} shows the heating rates computed 
from the initial compositions reproducing the 
solar $r$-process abundance ratios for $A \ge 70$ (purple), 
$A \ge 90$ (red), and $A \ge 120$ (blue).
The results are found to be similar 
as long as the minimal masses $A = 90 - 120$ are considered.
In this paper, we use the heating rate with $A \ge 90$,
being smaller than that adopted in TH13 
by a factor of about 3 at $t = 1-10$ days, 
and in good agreement with the mentioned nucleosynthetic results
\citep{metzger10,goriely11,korobkin12,grossman13}.

As in TH13, the effect of $\gamma$-ray transport is 
crudely taken into account 
by introducing a thermalization factor $\epsilon_{\rm therm}$
\citep{metzger10,korobkin12}.
A fraction $\epsilon_{\rm therm}$ of the decay energy $\dot{E}_{\rm decay}$
(in Figure \ref{fig:Qdot}) is assumed to be thermalized,
\ie $\dot{E}_{\rm rad} = \epsilon_{\rm therm} \dot{E}_{\rm decay}$,
where the energy $\dot{E}_{\rm rad}$ is immediately deposited.
We adopt $\epsilon_{\rm therm} = 0.5$.

For the elemental abundances in the ejecta, 
we assume homogeneous distribution 
with the solar abundance ratios as in TH13
(\ie detailed nucleosynthesis is not solved)
both for BH-NS and NS-NS mergers.
To be consistent with the assumption for the heating rate,
we include elements with $Z \ge 40$ (Zr and heavier).

A key ingredient of the simulations is calculation of the opacity.
The wavelength-dependent opacity is computed
by taking into account electron scattering, and bound-bound, 
bound-free, and free-free transitions.
Among these opacities, the bound-bound opacity is always dominant.
The code includes bound-bound opacities of 
almost all the $r$-process elements
from the VALD database \citep{piskunov95,ryabchikova97,kupka99,kupka00}.
As discussed in TH13, our line list 
includes the data for $r$-process elements only 
up to doubly ionized ions 
(there are no data for triply and more ionized ions in the VALD database).
As a result, the code cannot correctly evaluate the opacity 
(and gives an extremely low opacity) at the epoch of $t \lsim 1$ day, 
when the temperature is higher than about 10,000 K.
To avoid this artificially low opacity, we set 
a lower limit of the opacity of $\kappa_{\rm low} = 1 \ {\rm cm^2 \ g^{-1}}$,
and assume a gray opacity of $\kappa_{\rm low}$ when 
the computed Rosseland mean opacity is lower than this value.
For the most part of this paper, 
we do not discuss the emission at such early epochs,
which is affected by this assumption (see also Appendix B of TH13).

\begin{figure}
\begin{center}
\includegraphics[scale=1.3]{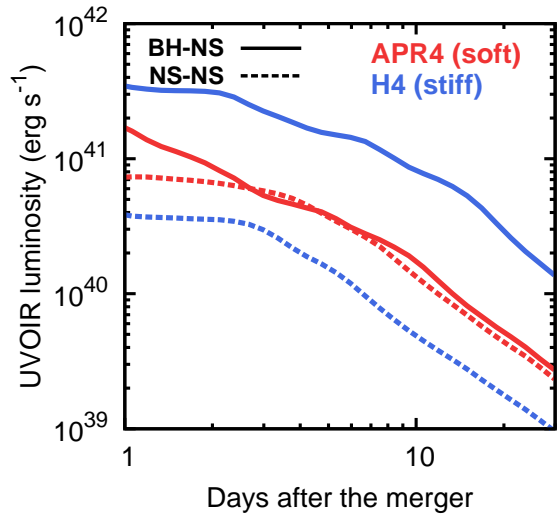}
\caption{
Bolometric light curves of the BH-NS (APR4Q3a75 and H4Q3a75, solid lines) 
and NS-NS merger models (APR4-1215 and H4-1215, dashed lines).
The luminosities are those averaged over all solid angles.
For the NS-NS merger models, 
the heating rate per ejecta mass is assumed to be the 
same with the BH-NS merger models (see the main text).
A stiffer EOS (blue lines) leads to a higher luminosity 
by the larger ejecta mass for BH-NS mergers
while it leads to a lower luminosity for NS-NS mergers.
}
\label{fig:Lbol_BHNS}
\end{center}
\end{figure}

\section{Results}
\label{sec:results}

\subsection{Dependence on the EOS and Comparison with NS-NS Mergers}

Figure \ref{fig:Lbol_BH} shows the computed light curves 
of the BH-NS merger models.
The luminosities are those averaged over all solid angles.
Because of the ejecta entirely made of $r$-process elements,
their opacities for the BH-NS mergers
reach $\kappa \sim 10 \ {\rm cm^2 \ g^{-1}}$ 
as in the NS-NS mergers (\citealt{kasen13,barnes13}; TH13).

The mass ejection in the BH-NS merger 
occurs dominantly by the tidal effect.
When a stiffer EOS, such as H4 or MS1, is adopted, 
the NS radius is larger
and the tidal disruption is more efficient.
Thus, the ejecta mass becomes larger with a stiffer EOS 
for a given mass ratio and BH spin \citep{kyutoku13}.
As a result, the models with stiffer EOSs are brighter
for a given mass ratio and BH spin,
provided that the heating rates are not dependent 
on the adopted EOSs.

The peak luminosity and the transition time to the 
declining phase approximately scale with 
$L \propto \Mej^{1/2}$ and $t \propto \Mej^{1/2}$, respectively, 
as expected by analytic formulae \citep{li98,metzger10}.
At declining phases, the photon diffusion is not important,
and the luminosity scales with $L \propto \Mej$ 
as long as a constant thermalization factor $\epsilon_{\rm therm}$
is adopted.

\begin{figure}
\begin{center}
\includegraphics[scale=1.3]{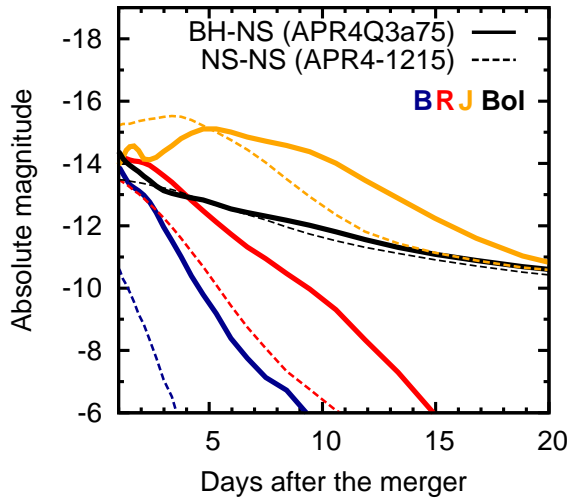}
\caption{
The $B$, $R$, $J$-band, and bolometric light curves of 
the BH-NS merger model APR4Q3a75 (solid) and 
the NS-NS merger model APR4-1215 (dashed).
Although the bolometric light curves of these two models
(with similar ejecta masses, $\Mej \sim 0.01 \Msun$)
are similar, the light curves of the BH-NS merger model 
in the optical bands ($B$ and $R$ bands in this figure) 
are brighter than those of the NS-NS merger model.
}
\label{fig:mag_BHNS}
\end{center}
\end{figure}

Figure \ref{fig:Lbol_BHNS} shows the bolometric light curves
of the BH-NS merger models (solid lines) 
compared with those of the NS-NS merger models (dashed lines). 
For the NS-NS merger models, 
the gravitational masses of two NSs are $1.2 \Msun$ and $1.5 \Msun$
\citep{hotokezaka13}.
The light curves of these NS-NS merger models have been 
shown in TH13, but for ease of comparison, 
we show the light curves computed with the same heating rate 
taken for the BH-NS merger models.

For the mass ratio ($Q=3$) and BH spin parameter ($\chi = 0.75$) 
adopted in our models, 
mass ejection from BH-NS mergers tends to be more efficient than 
that from NS-NS mergers.
The light curves of such BH-NS merger models (solid lines) 
are more luminous than those of NS-NS merger models 
as long as the same heating rate is assumed
\footnote{Since the dominant mechanism of mass ejection in
NS-NS mergers can be shock heating 
(especially when the mass ratio of the two NSs is close to unity,
\citealt{hotokezaka13}), $Y_\mathrm{e}$ in the ejecta 
can be quite different between NS-NS and BH-NS mergers.
As discussed by \citet{grossman13}, 
such a difference can affect the heating rate.}.
Among the NS-NS merger models shown in TH13, 
the model with the APR4 EOS (red dashed line in Figure \ref{fig:Lbol_BHNS}) 
gives the highest luminosity.
The luminosities of the BH-NS merger models H4Q3a75
and MS1Q3a75 are higher than that of the NS-NS merger model with the APR4 EOS
by a factor of $\sim 5$.
As already discussed in \citet{kyutoku13}, 
the dependences on EOSs are opposite between
BH-NS and NS-NS mergers;
a stiffer EOS leads to brighter light curves in BH-NS mergers 
while it results in fainter light curves in NS-NS mergers.

\begin{figure}
\begin{center}
\includegraphics[scale=1.3]{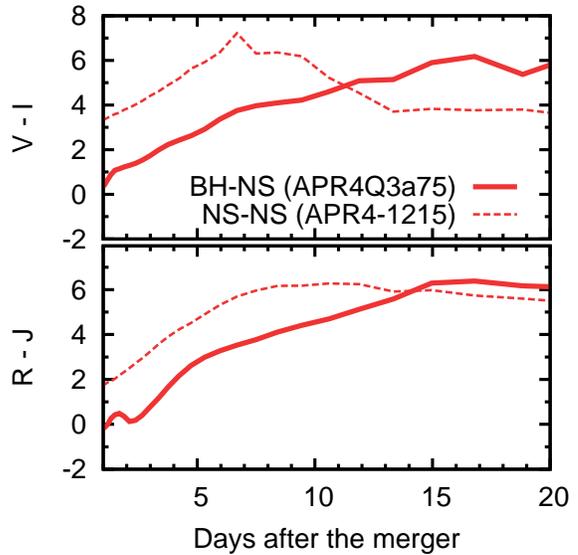}
\caption{
Color evolutions (Top: $V-I$ color, Bottom: $R-J$ color)
of the BH-NS merger model APR4Q3a75 (solid) and 
the NS-NS merger model APR4-1215 (dashed).
The BH-NS merger model has bluer colors in the first 10 days.
}
\label{fig:color_BHNS}
\end{center}
\end{figure}

\begin{figure}
\begin{center}
\includegraphics[scale=1.2]{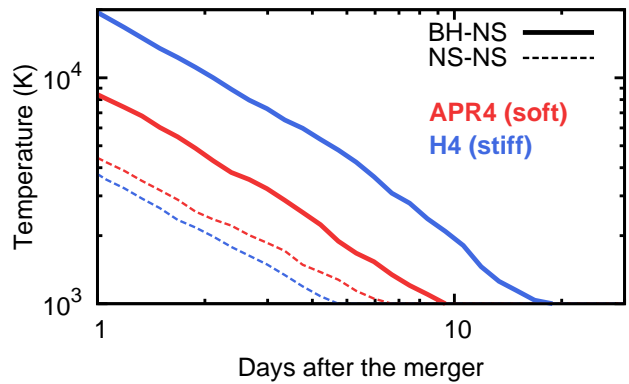}
\caption{
Temperatures of the BH-NS merger models (solid lines) and 
the NS-NS merger models (dashed lines)
at $v = 0.25 c$ (near the characteristic velocities of the models).
Red and blue lines show the models with
the soft (APR4) and stiff (H4) EOSs, respectively.
The BH-NS merger models tend to have a higher temperature,
which results in a bluer emission (Figures \ref{fig:mag_BHNS}
and \ref{fig:color_BHNS}).
}
\label{fig:temp}
\end{center}
\end{figure}

Interestingly, even with the similar ejecta mass, 
the behaviors of multi-band light curves can be different 
between BH-NS and NS-NS mergers.
Figure \ref{fig:mag_BHNS} shows the multi-band light curves of 
the BH-NS merger model APR4Q3a75 (solid line) and the NS-NS merger model 
APR4-1215 (dashed line).
Although these two models have similar bolometric luminosities
(Figure \ref{fig:Lbol_BHNS})
\footnote{The light curve of BH-NS merger model APR4Q3a75 
has a slightly higher peak luminosity 
at earlier epochs than that of the NS-NS merger model APR4-1215.
As already discussed by \citet{kyutoku13},
this is due to a short diffusion length/timescale 
for the BH-NS merger ejecta.},
the light curves of the BH-NS merger model 
in the optical bands ($UBVRI$) 
are brighter than those of the NS-NS merger model.
This difference is clearly shown in the color evolutions
shown in Figure \ref{fig:color_BHNS}.
The $V-I$ and $R-J$ colors of the BH-NS merger are bluer 
than those of the NS-NS merger by 2 dex during the first 10 days.

The difference in the color evolutions mainly results 
from the ejecta geometries.
Figure \ref{fig:temp} shows 
the temperatures at $v = 0.25 c$, 
near the characteristic velocities $v_{\rm ch}$ of the models.
The temperatures for the BH-NS merger models 
are systematically higher than those of the NS-NS merger models.
It is emphasized that there is such a difference even 
for similar ejecta masses;
the models APR4Q3a75 and APR4-1215 have similar 
$\Mej (\sim 10^{-2} \Msun$), but their temperatures are different 
by a factor of about 2.
In the model APR4Q3a75, the mass ejection is confined 
in a small 
solid angle near the equatorial plane.
Thus, for a given ejecta mass, the ejecta matter 
of the BH-NS merger model
has a smaller volume than that of the NS-NS merger model.
Since a similar radiation energy is deposited in the small volume,
the temperature of the BH-NS merger model becomes higher.
As a result, the emission from BH-NS merger ejecta tends to be bluer
when the mass ejection is confined in a small solid angle.
Implications of this trend are discussed in Section \ref{sec:implications}.

\begin{figure}
\begin{center}
\includegraphics[scale=1.3]{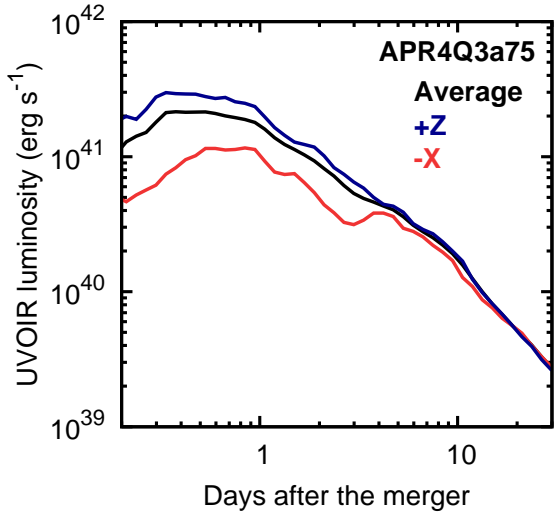}
\caption{
Bolometric light curves of the BH-NS merger model APR4Q3a75 
viewed from different lines of sight
(in isotropic luminosity).
The light curve observed from the direction of
mass ejection ($-x$ in the right panel of Figure \ref{fig:models})
is fainter than those viewed from the other directions.
Note that our simulations assume
the gray opacity of $\kappa = 1\ {\rm cm^2 \ g^{-1}}$
at $t \lsim 1$ day, since our line list for bound-bound transitions
is not applicable
at such early epochs.
}
\label{fig:Lbol_angle}
\end{center}
\end{figure}

\begin{figure*}
\begin{center}
\begin{tabular}{cc}
\includegraphics[scale=1.05]{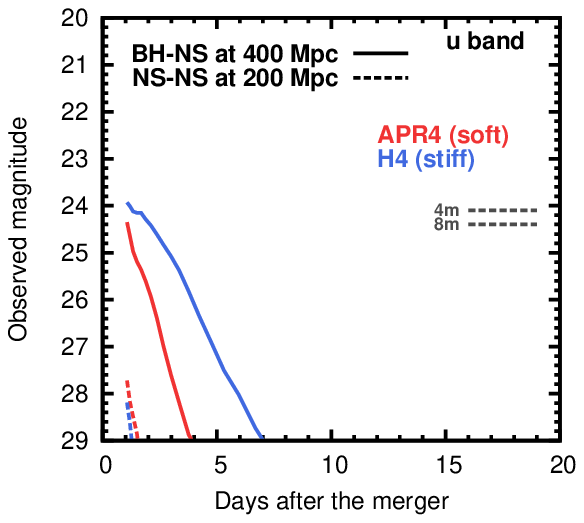} &
\includegraphics[scale=1.05]{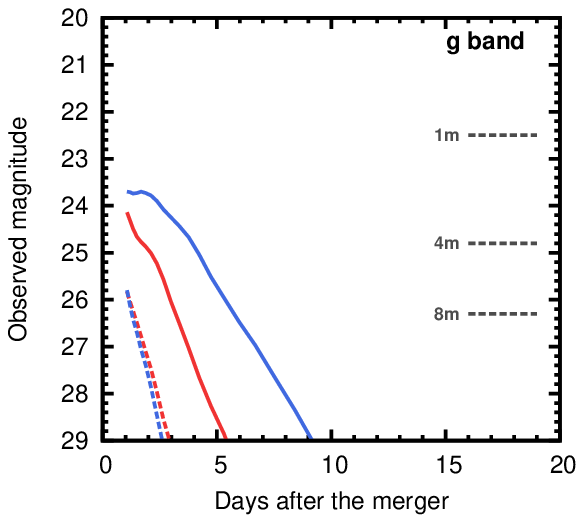} \\
\includegraphics[scale=1.05]{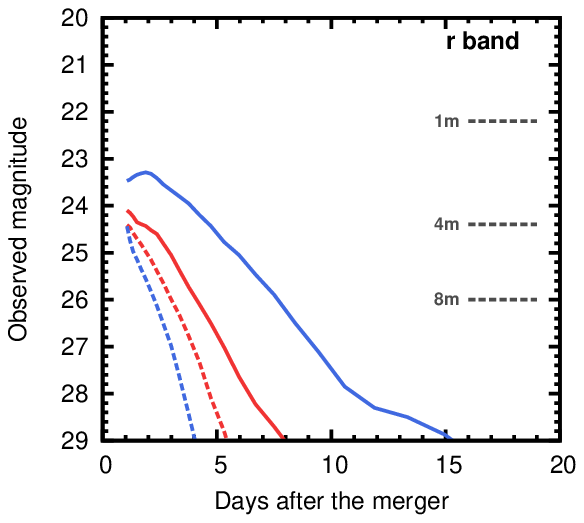} &
\includegraphics[scale=1.05]{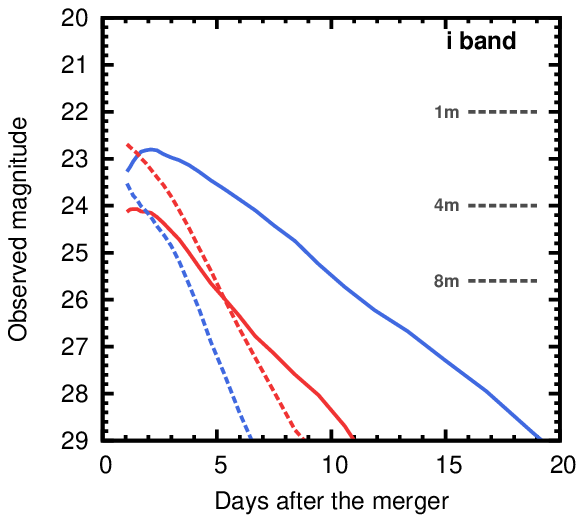} \\
\includegraphics[scale=1.05]{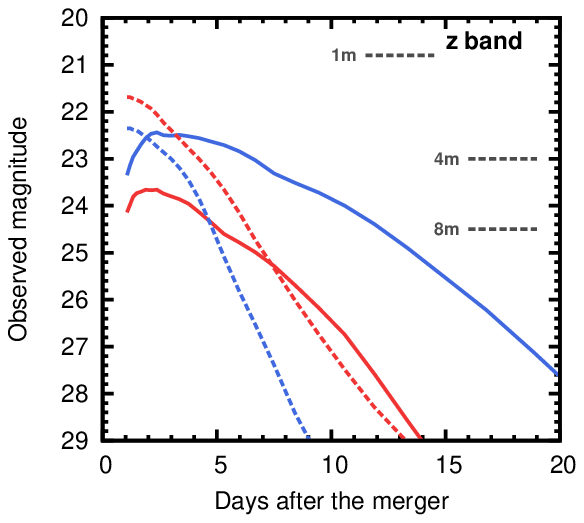} & 
\includegraphics[scale=1.05]{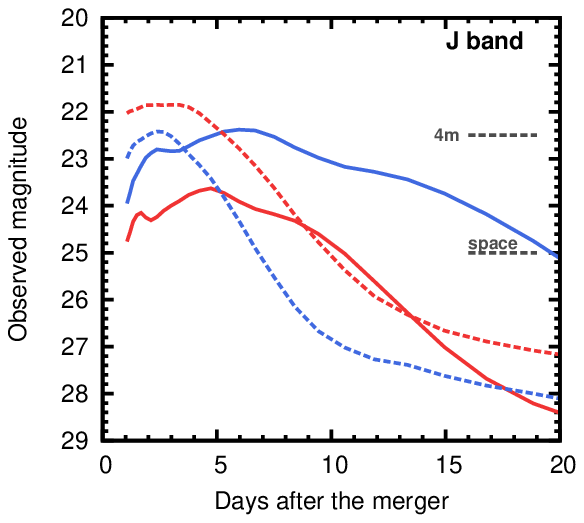} \\
\includegraphics[scale=1.05]{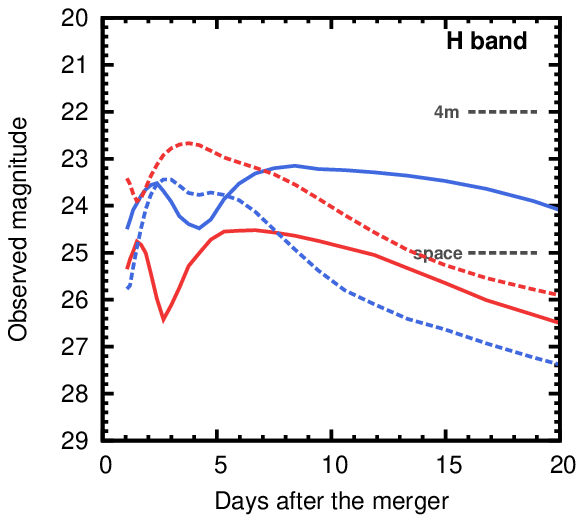} &
\includegraphics[scale=1.05]{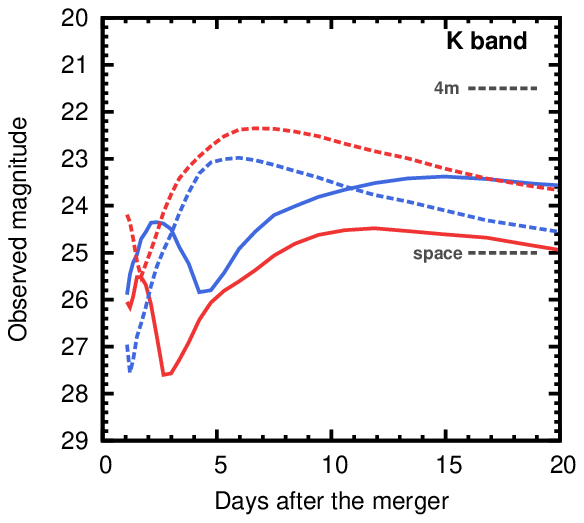}  
\end{tabular}
\caption{
Expected observed $ugrizJHK$-band light curves 
(in AB magnitudes) for the BH-NS merger models
APR4Q3a75 (red solid lines) and H4Q3a75 (blue solid lines) 
and the NS-NS merger models
APR4-1215 (red dashed lines) and H4-1215 (blue dashed lines).
The light curves are those averaged over all solid angles.
The distances to the events are set to be 400 Mpc
(BH-NS) and 200 Mpc (NS-NS).
$K$ correction is taken into account.
Horizontal lines show typical limiting magnitudes for 
wide-field telescopes ($5 \sigma$ with 10 min exposure).
For optical wavelengths ($ugriz$ bands), 
``1 m'', ``4 m'', and ``8 m'' limits are 
taken or deduced from those of PTF \citep{law09}, 
CFHT/Megacam, and Subaru/HSC \citep{miyazaki06}, respectively.
For NIR wavelengths ($JHK$ bands), ``4 m'' and ``space'' limits 
are taken or deduced from those of Vista/VIRCAM
and the planned limits of WFIRST 
\citep{green12} and WISH \citep{yamada12}, respectively.
}
\label{fig:mag_d400}
\end{center}
\end{figure*}

\begin{figure}
\begin{center}
\includegraphics[scale=1.5]{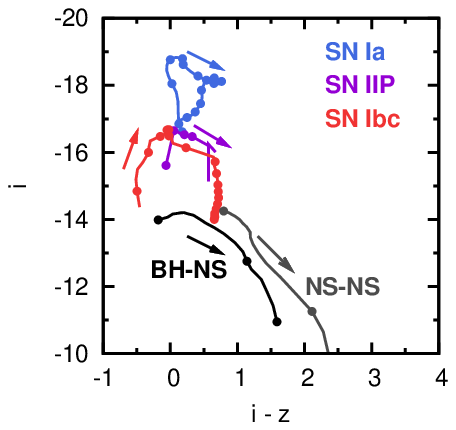}
\includegraphics[scale=1.5]{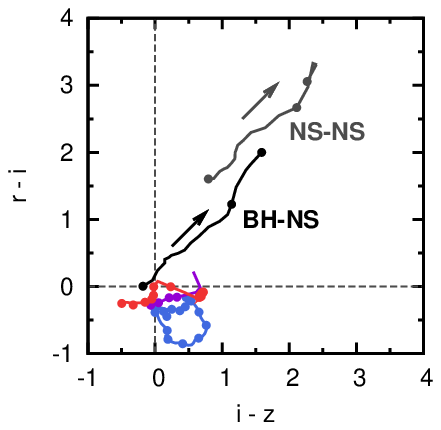}
\caption{
Color-magnitude (Top) and color-color (Bottom) diagrams
for the BH-NS merger model APR4Q3a75 (black)
and the NS-NS merger model APR4-1215 (gray).
These models are compared with Type Ia (blue),
IIP (purple), and Ibc (red) supernovae.
The BH-NS and NS-NS mergers have fainter absolute magnitudes,
and redder colors than supernovae.
The arrows show the direction of time evolution,
and dots for each model are shown with 5-day intervals.
For supernovae, we use the spectral templates by \citet{nugent02}.
All the magnitudes are in AB magnitudes and 
in the rest frame (\ie no $K$ correction).
Dashed lines in the bottom panel show the positions of $r-i = 0$
and $i-z=0$.
}
\label{fig:HR}
\end{center}
\end{figure}

\subsection{Viewing Angle Effects}

When the mass ejection from BH-NS mergers is highly asymmetric,
the behaviors of the light curves are expected 
to depend on the line of sight of observers \citep{kyutoku13}.
Figure \ref{fig:Lbol_angle} shows the bolometric 
light curves of the model APR4Q3a75 viewed from different angles.
In this plot, we include the light curves at $t < 1$ day
since the viewing angle effects are most important 
at the earliest epochs.
It should be cautioned that our simulations assume
the gray opacity of $\kappa = 1\ {\rm cm^2 \ g^{-1}}$
at $t \lsim 1$ day, since our line list for bound-bound transitions
is not applicable
at early epochs when the temperature is $T \gsim 10,000$ K (see TH13).
Nevertheless, relative behaviors 
of the light curves may be worth discussing.

We find that the emission viewed from 
the direction of mass ejection 
($-x$ direction in the right panel of Figure \ref{fig:models})
is fainter than those from the other viewing angles.
This is because the diffusion path, and hence the diffusion timescale, 
are longer when the BH-NS merger is viewed from the direction 
of mass ejection.
The viewing angle effect will be important only at the first five days,
when the ejecta are opaque and the photon diffusion is important.
Since the solid angle of mass ejection 
(\ie the probability to observe BH-NS events from the direction
of the mass ejection) is small,
we do not expect that the viewing angle effects have 
a big impact on follow-up observations.

\section{Implications for Follow-up Observations}
\label{sec:implications}

Based on the results of our simulations, 
we discuss a strategy for follow-up observations
of EM counterparts after the GW detection from BH-NS mergers.
Figure \ref{fig:mag_d400} shows 
expected multi-band light curves 
(in AB magnitudes for SDSS $ugriz$ filters and NIR $JHK$ filters)
for the BH-NS merger models
APR4Q3a75 (red solid lines) and H4Q3a75 (blue solid lines) and 
the NS-NS merger models APR4-1215 (red dashed lines) 
and H4-1215 (blue dashed lines).
Since a typical distance to BH-NS merger events
is expected to be larger than that to NS-NS merger events \citep{abadie10rate}, 
the distance to the BH-NS merger events is assumed to be 400 Mpc, 
while that to the NS-NS merger events is 200 Mpc.

For the mass ratio ($Q=3$) and BH spin parameter ($\chi = 0.75$) 
aligned with the binary orbital angular momentum,
the BH-NS merger models have relatively large ejecta masses.
In such cases, thanks to the intrinsically higher 
luminosities and bluer colors,
observed magnitudes of BH-NS mergers can be comparable to
or even brighter than those of NS-NS mergers,
compensating the larger distance.
As discussed in Section \ref{sec:results},
the dependence on the EOS is opposite for BH-NS and NS-NS mergers.
The mass ejection for BH-NS mergers is more efficient with a stiff EOS
\citep{kyutoku13}
while that for NS-NS mergers is more efficient with a soft EOS
\citep{hotokezaka13,bauswein13}.
Thus, if a stiff EOS, such as H4 and MS1, is the case, 
radioactively powered emission from 
BH-NS mergers can be more easily detected than that from NS-NS mergers 
(see blue lines in Figure \ref{fig:mag_d400}).

When the mass ejection for BH-NS mergers is 
efficient as in the models adopted in this paper,
a similar follow-up strategy can be applied 
both for the BH-NS and NS-NS merger events.
As already discussed by \citet{barnes13} and TH13,
observations in the red optical and NIR wavelengths are the most efficient.
In optical wavelengths,
observations with wide-field 4m- and 8m-class telescopes are necessary.
Such facilities include 
3.6m Canada-France-Hawaii Telescope (CFHT)/Megacam 
(3.6 deg$^2$ field of view, FOV),
the Blanco 4m telescope/DECAM (4.0 deg$^2$ FOV), 
8.2m Subaru/Hyper Suprime Cam \citep[HSC,][1.77 deg$^2$ FOV]{miyazaki06}, 
and 8.4m Large Synoptic Survey Telescope 
(LSST, \citealt{ivezic08,lsst09}, 9.6 deg$^2$ FOV).
In NIR wavelengths, observations with wide-field space telescopes,
such as The Wide-Field Infrared Survey Telescope 
(WFIRST, \citealt{green12}, 0.375 deg$^2$ FOV)
and Wide-field Imaging Surveyor for High-redshift 
(WISH, \citealt{yamada12}, 0.28 deg$^2$ FOV)
will be important.

Since BH-NS and NS-NS mergers are rare events,
follow-up observations for the EM counterparts of the GW sources 
may discover more supernovae, 
which occur by chance within the localization area of GW sources.
Thus, classification of transient objects is extremely important.
Since the timescale of the light curve evolution for
BH-NS and NS-NS merger events
is much faster than that for supernovae, 
multiple visits within 5-10 days are the most effective way to classify 
the transient objects as BH-NS or NS-NS merger events.
However, even with observations at a single or a few epochs,
a classification may be possible.
Figure \ref{fig:HR} shows the color-magnitude (top) 
and color-color (bottom) diagrams.
These diagrams show that
the radioactively powered emission from
BH-NS and NS-NS mergers are fainter and redder than that from supernovae.
Only at the brightest phase, the BH-NS merger models can have 
a similar color to that of supernovae.
Thus, even without detailed light curves,
we may be able to distinguish BH-NS and NS-NS mergers from supernovae
with multi-band observations.

Color information can be used to effectively pick up the candidates.
However, 
to conclusively identify the transient objects as BH-NS or NS-NS mergers,
spectroscopic observations are eventually necessary.
If extremely broad-line, smeared-out spectra are obtained
(see Figure 6 of TH13),
such an object is likely to be the counterpart of a GW source.

Interestingly, we might be able to even distinguish 
BH-NS mergers from NS-NS mergers
by the radioactively powered emission.
When the mass ejection from a BH-NS merger
is confined in a small solid angle,
the emission from the ejecta can have bluer colors than 
those for NS-NS mergers
(Figures \ref{fig:color_BHNS} and \ref{fig:HR}).
In such a case, the emissions
for BH-NS and NS-NS merger models 
occupy different regions in a color-color diagram (Figure \ref{fig:HR}).
In order to find the general emission properties of BH-NS merger ejecta,
we have to study a wide variety of possible models, such as  
models with different BH to NS mass ratios, 
with different BH spin parameters, and with non-aligned BH spins. 
Contributions from the $r$-processed ejecta from a
BH-accretion torus, expected to form after the first dynamical matter ejection,
should also be taken into account in the future study
\citep{surman08,wanajo12,fernandez13}. 
Nevertheless, we emphasize that, in addition to the GW observations
\citep[see \eg][]{hannam13},
multi-band optical and NIR observations of radioactively powered emission 
may also provide independent information on 
the progenitors of GW sources (see also \citealt{hotokezaka13b}).

\section{Conclusions}
\label{sec:conclusions}

We have performed three-dimensional, 
time-dependent, multi-frequency Monte-Carlo radiative transfer 
simulations for radioactively powered emission from BH-NS mergers
by taking into account the wavelength-dependent opacities of 
$r$-process elements.
We showed that, for the BH to NS mass ratio of $Q=3$ and 
BH spin parameter of $\chi = 0.75$
aligned with the orbital angular momentum, 
radioactively powered emission from BH-NS mergers can be 
more luminous than that from NS-NS mergers.
In such cases, the observed brightness 
of BH-NS mergers can be comparable to or even higher
than that of NS-NS mergers,
compensating expected typical larger distances to BH-NS mergers.
Then, a similar observational strategy to identify EM counterparts 
works both for the BH-NS and NS-NS merger events.
Observations at the red edge of optical and NIR wavelengths 
are most efficient.
If a stiff EOS is the case, the EM counterparts of GW sources 
can be more easily detected for BH-NS mergers than for NS-NS mergers.

When the mass ejection from a BH-NS merger 
is confined in a small solid angle,
a large radioactive energy is deposited to the small volume, 
which makes the ejecta temperature 
higher than that for an NS-NS merger.
As a result, the emission from BH-NS mergers
can be bluer than that from NS-NS mergers.
Thanks to these properties, 
we might be able to distinguish BH-NS events from NS-NS merger events
by multi-band observations of radioactively powered emission.
Although the general emission properties of the BH-NS merger ejecta
are still unknown, owing to, \eg 
unknown mass functions and spin parameters of BHs,
our results demonstrate that 
EM observations can potentially provide
independent information on the progenitors of GW sources
and the nature of compact binary coalescences.

\acknowledgments

This work was in part developed during the long-term workshop on 
{\em Gravitational Waves and Numerical Relativity} held at the Yukawa
Institute for Theoretical Physics, 
Kyoto University in May and June 2013.
We have made use of NIST database for atomic data, 
and VALD database \citep{piskunov95,ryabchikova97,kupka99,kupka00}
for line lists.
Atomic data compiled in the DREAM data base \citep{biemont99} 
were extracted via VALD.
The numerical simulations presented in this paper 
were carried out with Cray XC30 at Center for Computational Astrophysics, 
National Astronomical Observatory of Japan.
This research has been supported 
by the Grant-in-Aid for Scientific Research of the 
Japan Society for the Promotion of Science 
(JSPS, 23224004, 23740160, 24244028, 24740117, 24740163)
and Grant-in-Aid for Scientific Research on Innovative Areas
of the Ministry of Education, Culture, Sports, Science and Technology 
(MEXT, 25103510, 25103512, 25103515, 25105508).
Kenta Hotokezaka is supported by JSPS fellowship Grant Number 24-1772.
Koutarou Kyutoku is supported by JSPS Postdoctoral Fellowships 
for Research Abroad.

\end{document}